\documentclass[letter]{aa}  
\usepackage{xcolor}
\usepackage{layouts}
\usepackage{graphicx}
\usepackage{todonotes}
\usepackage{rotating} 
\usepackage{caption}
\usepackage{todonotes}
\usepackage{lipsum}

\usepackage{graphicx}
\usepackage{placeins}
\usepackage[breaklinks=true]{hyperref}
\usepackage{natbib,twoopt}
\bibpunct{(}{)}{;}{a}{}{,}             
\makeatletter
  \newcommandtwoopt{\citeads}[3][][]{\href{http://adsabs.harvard.edu/abs/#3}%
    {\def\hyper@linkstart##1##2{}%
     \let\hyper@linkend\@empty\citealp[#1][#2]{#3}}}
  \newcommandtwoopt{\citepads}[3][][]{\href{http://adsabs.harvard.edu/abs/#3}%
    {\def\hyper@linkstart##1##2{}%
     \let\hyper@linkend\@empty\citep[#1][#2]{#3}}}
  \newcommandtwoopt{\citetads}[3][][]{\href{http://adsabs.harvard.edu/abs/#3}%
    {\def\hyper@linkstart##1##2{}%
     \let\hyper@linkend\@empty\citet[#1][#2]{#3}}}
  \newcommandtwoopt{\citeyearads}[3][][]%
    {\href{http://adsabs.harvard.edu/abs/#3}
    {\def\hyper@linkstart##1##2{}%
     \let\hyper@linkend\@empty\citeyear[#1][#2]{#3}}}
\makeatother

\usepackage{graphicx}
\usepackage{booktabs}
\usepackage{lipsum}
\usepackage{txfonts}
\usepackage{hyperref}
\usepackage{tabularx}
\usepackage{upgreek}

\graphicspath{{Figures/}} 

\begin{document} 
\abstract{

 We report the discovery of complex flaring activity from the galactic nucleus hosting the five-year-old tidal disruption event \object{eRASSt~J234402.9$-$352640} (J2344). With \textit{Einstein Probe} and \textit{XMM-Newton} observations, we detected highly structured soft X-ray variability. Through temporal decomposition of the \textit{XMM-Newton} light curve and time-resolved spectral analysis, we identified broad, thermal flares recurring every $\sim12$~hours and lasting $\sim2$~hours, consistent with quasi-periodic eruptions (QPEs). Remarkably, these QPEs are accompanied by an unprecedented crest of hotter shorter flares, each lasting between 5 and 30~minutes. These flares are predominantly found in the rising phases of the QPEs, although they also appear throughout the quiescence. These findings establish J2344 as a new member of the QPE emitter population and uncover a previously unobserved phenomenology that challenges current models of QPEs. In this letter we present the phenomenological properties of this unique source and discuss possible interpretations within the framework of extreme mass ratio inspirals.
}
\vspace{-0.1in}

\vspace{-0.1in}

\authorrunning{P. Baldini et al}
\title{Discovery of crested quasi-periodic eruptions following the most luminous SRG/eROSITA tidal disruption event}

\author{P. Baldini\inst{1}
\and A. Rau\inst{1}
\and A. Merloni\inst{1}
\and B. Trakhtenbrot\inst{1,2,3}
\and R. Arcodia\inst{4}
\and M. Giustini\inst{5}
\and G. Miniutti\inst{5}
\and S. J. Brennan\inst{1} 
\and \\ M. Freyberg\inst{1}
\and P. Sánchez-Sáez \inst{6}
\and I. Grotova\inst{1}
\and Z. Liu\inst{7}
\and T. Lian\inst{1}
\and K. Nandra\inst{1}
}

\institute{Max-Planck-Institut für extraterrestrische Physik, Gießenbachstraße 1, 85748 Garching, Germany: \href{mailto:baldini@mpe.mpg.de}{baldini@mpe.mpg.de}
\and Excellence Cluster ORIGINS, Boltzmannsstraße 2, 85748, Garching, Germany
\and School of Physics and Astronomy, Tel Aviv University, Tel Aviv 69978, Israel
\and Kavli Institute for Astrophysics and Space Research, Massachusetts Institute of Technology, Cambridge, MA 02139, USA
\and Centro de Astrobiologia (CAB), CSIC-INTA, Camino Bajo del Castillo s/n, ESAC, 28692, Villanueva de la Cañada,
Madrid, Spain
\and European Southern Observatory, Karl-Schwarzschild-Strasse 2, 85748 Garching bei München, Germany
\and Centre for Astrophysics Research, University of Hertfordshire, College Lane, Hatfield AL10 9AB, UK
}

\maketitle
\vspace{-0.1in}

\section{Introduction}
\vspace{-0.1in}

Quasi-periodic eruptions (QPEs) are soft X-ray flares from supermassive black holes (SMBHs) in galactic nuclei that repeat on timescales between a few hours and a few days (\citealp{miniutti2019nine, giustini2020quasi, arcodia2021, chakraborty2021, quintin2023, bykov2023,arcodia2024a, nicholl2024qpe, chakraborty2025, hernandezgarcia2025, arcodia2025}). Their spectra are well modeled with evolving blackbodies, with temperatures of 50--250\,eV, and bolometric luminosities of $10^{42-43}$ erg/s.

The absence of correlated variability outside the X-ray band (e.g., \citealp{giustini2024, wevers2025, Goodwin2025}) complicates the efforts to understand their nature. However, a crucial piece of information comes from the increasingly secure association of QPEs with tidal disruption events (TDEs; see \citealp{jonker2021tidal} for an extensive review). The first QPEs were  discovered in the late-time light curve of the repeating X-ray TDE in the nucleus of GSN 069 (\citealp{shu2018, miniutti2019nine, miniutti2023, sheng2021}). Of the remaining ten known QPEs, discovered either through archival searches or serendipitously, 
five have been associated with TDEs or general episodic accretion events (\citealp{chakraborty2021, quintin2023, bykov2023, nicholl2024qpe, chakraborty2025, hernandezgarcia2025, zhu2025ansky}). Further observational evidence connecting TDEs and QPEs was reported in \citealp{sheng2021, arcodia2024a,gilbert2024,wevers2024, kosec2025}.

The association of TDEs and QPEs is corroborated by theoretical models. The most supported scenarios involve the presence of a stellar-mass object orbiting the SMBH, in so-called extreme mass ratio inspirals (EMRIs; \citealp{amaroseaone2018}). In this framework, if an accretion disk forms on a plane different from the orbital plane of the stellar-mass object, QPEs will be produced as the disk and the object collide with each other (e.g., \citealp{dai2010, xian2021, franchini2023, linial2023, tagawa2023, zhou2024, vurm2025}). If the stellar-mass object is a star, each passage will strip its envelope, which will participate in the collisions, giving rise to the complex light curves that are sometimes observed in QPEs (\citealp{yao2025, linial2025, hernandezgarcia2025b}). Even though this family of models has passed more tests than those invoking periodic processes in the accretion flow (e.g., \citealp{raj2021, pan2022disk, sniegowska2023}), a consensus has yet to be reached (e.g., \citealp{mummery2025}). In this letter, we report the discovery of QPEs five years after the TDE eRASSt J234402.9-352640 (hereafter J2344; \citealp{homan2023, goodwin2024}, \citealp{Malyali2025}), strengthening the TDE-QPE connection, while simultaneously challenging our understanding of the nature of these enigmatic events.

\vspace{-0.1in}
\section{Discovery and analysis}
\vspace{-0.05in}
\begin{figure*}[t]
    \centering
    \includegraphics[width=1\textwidth]{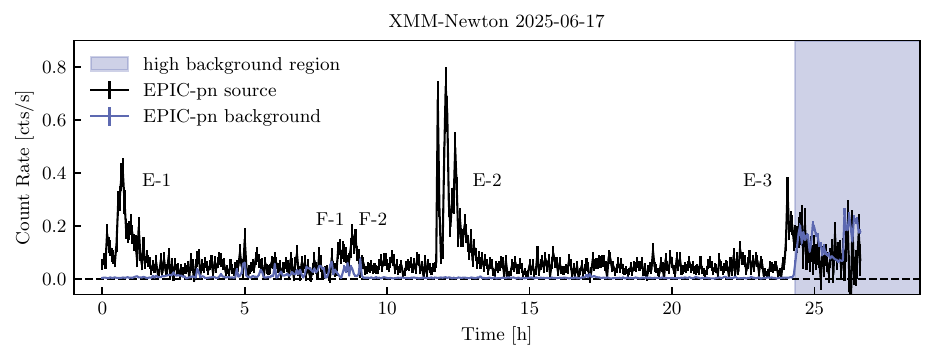}
    \vspace{-0.3in}
    \caption{XMM-Newton EPIC-pn 0.2--10 keV light curve of J2344. The source count rates are shown in black, while the background is plotted in blue. Due to the selected observing strategy, only short snapshots spaced by several hours are available for EP.}
    \label{fig:lctot}
    \vspace{-0.15in}

\end{figure*}

We observed J2344 with the Follow-up X-ray Telescope \citep[FXT;][]{chen2025} on board the Einstein Probe (EP) mission \citep{yuan2025} on 2024 October 22 for 6$\times$1.2 ks snapshots over a baseline of 22 hours. The combined light curve of the two instrument modules, FXTA and FXTB, showed significant variability (see Appendices \ref{a:time} and \ref{a:complc}). The source was observed again for 12\,ks over 6 hours on 2025 June 02  and, after confirming the presence of repeating flares, a 99\,ks XMM-Newton observation was performed on 2025 June 17. The XMM-Newton light curve, shown in Fig.~\ref{fig:lctot}, clearly reveals two bright flares (hereafter E-1 and E-2) with superimposed short-time variability. Additionally, a potential third flare (E-3) was detected at the end of the observation, shortly before the background started to dominate. Two narrower and fainter flares, F1 and F2, were found between E-1 and E-2. We then reobserved J2344 with EP on 2025 July 28, and confirmed the flaring activity (Appendix \ref{a:complc}).

We modeled the XMM-Newton light curve with a combination of Gaussians (see Fig. \ref{fig:totplot}, and details in Appendix~\ref{a:time}). E-1 and E-2 can each be modeled with a broad Gaussian with a similar width and amplitude, and  respectively two and three narrower Gaussians to account for the narrow flares  (see Table \ref{tab:time}). E-3 is consistent with the same broad Gaussian profiles as E-1 and E-2 (although the background is dominating), plus one narrow feature. F1 and F2 can be modeled as two narrow Gaussians.

We performed time-resolved spectral analysis along these  features within a Bayesian framework (see Appendix~\ref{a:xrspec} for details). For E-1, E-2, and E-3 we extracted spectra corresponding to the narrow flares, as well as those periods in between, to describe the spectral evolution of both the broad and narrow modulations. For F-1 and F-2, we extracted spectra to account for each individual flare. For all segments, the source spectra can be well modeled as a blackbody in addition to the quiescent emission, for which the spectrum is shown and discussed in Appendix \ref{a:xrspec}. We note, however, that although not strictly required, the light curve decomposition would suggest modeling the narrow flares as an additional blackbody component. We report the details and results of this approach in Appendix \ref{a:xrspec}. In the top two panels of Fig. \ref{fig:totplot}, we show the evolution of the model-derived blackbody temperatures (T) and bolometric luminosities ($L_{\rm bol}$). The same results are quantified in Table \ref{tab:spec}.  
\vspace{-0.1in}
\section{Results}
\begin{figure*}[t]
    \centering
    \includegraphics[width=0.95\textwidth]{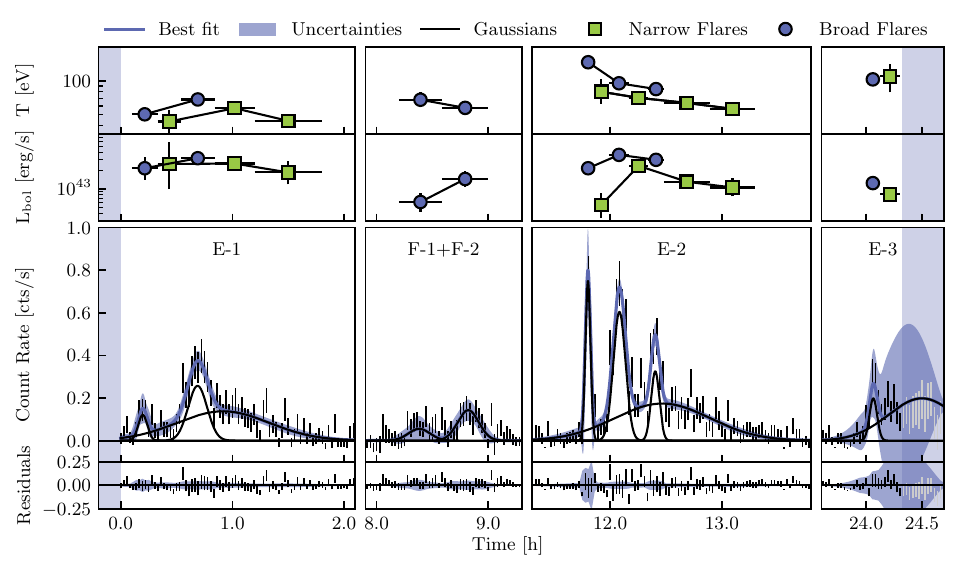}
    \vspace{-0.2in}
    \caption{Best fit of the XMM-Newton flare profiles. We do not find significant evidence of asymmetry; therefore, we use Gaussian functions. In the top two panels, we show the temperature and bolometric luminosity evolution along the flares, derived from time-resolved spectral analysis. The blue points correspond to data extracted during the short-time flares, while the green points trace the evolution of the broad underlying flares.}
    \label{fig:totplot}
    \vspace{-0.15in}
\end{figure*}

\begin{figure}[!htb]
     \centering
    \includegraphics[width=0.95\columnwidth]{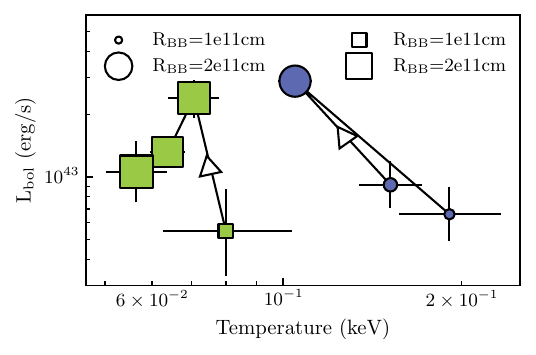}
     \vspace{-0.2in}
    \caption{Spectral evolution of the broad component (green) and of the three stacked narrow flares of E-2 (blue). While the broad component evolves counterclockwise, consistent with typical QPE behavior, the narrow flares show the opposite behavior.}
    \label{fig:hysteresis_m}
    \vspace{-0.1in}
\end{figure}

\subsection{The broad flares are QPEs}

We interpret the broad components E-1, E-2, and E-3 as QPEs. The quasi-periodicity is supported by the comparable Gaussian profiles, with an average $\sigma=0.43 \pm 0.05$ hours, and separated on average by $12\pm1$ hours ($11.7\pm0.1$ hours when the less constrained E-3 flares are removed). The second piece of evidence can be found instead in the time-resolved spectra. To date, all known bona fide QPEs have shown a characteristic hysteresis cycle in the L$_{\rm bol}$ versus T plane along a single eruption (e.g., \citealp{miniutti2023,arcodia2025}), a behavior also observed in the radiation-instability driven heartbeats of the black hole X-ray binary IGR J17091-3624 (\citealp{altamirano2011}). 
This characteristic cycle is consistent with the thermal emission being produced by an expanding and later contracting emitting region. In Fig. \ref{fig:hysteresis_m}, we plot the evolution of the best sampled broad flare (E-2) in the L$_{\rm bol}$ versus T plane, showing the derived radius of the emitting region as different-sized markers. The broad component of J2344 is consistent with a counterclockwise cycle, in agreement with other QPEs. We discuss the caveats of this analysis in Appendix~\ref{a:xrspec}.  

\begin{figure}[ht]
    \centering
    \includegraphics[width=\columnwidth]{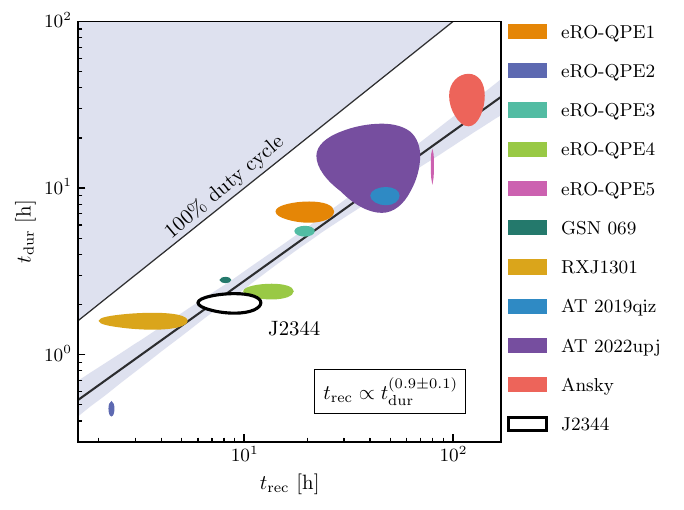}
     \vspace{-0.3in}
    \caption{Duration vs. recurrence time relation for all known QPEs. The black contours represent the newly added J2344. We conservatively estimate a lower limit on the recurrence time of $11.7/2\,$h. This accounts for the possibility of missing an intermediate broad flare, due to modulations of the QPE amplitude (e.g., \citealp{giustini2024}). The top left shaded area  corresponds to a duration time greater than or equal to the recurrence time.}
    \label{fig:durrec}
    \vspace{-0.2in}
\end{figure}

Quasi-periodic eruptions have also been observed to obey a tight and possibly linear recurrence versus duration time relation (see, e.g., \citealp{arcodia2025}). In Fig. \ref{fig:durrec} we show J2344 with the rest of the QPE emitters in such a plane. We recomputed the duration times as the width at $1/e^3$ of the eruption peak flux. J2344 is in agreement with the known trend. We re-derive the analytical relation as $t_{\rm dur} = (0.33\pm0.04) {t_{\rm rec}}^{0.9\pm 0.1}$, with an intrinsic scatter of $0.18\pm0.06$. If QPEs are due to EMRIs, the inferred $t_{\rm rec}$ implies an orbital radius of 12--32 gravitational radii (R$_g$), assuming a SMBH mass of log$(M_{BH})=7.2 \pm 0.2$ (\citealp{Malyali2025}, as revised with respect to the value of log$(M_{BH})\sim7.8$ reported in \citealp{homan2023}). We note, however, that the timing properties of J2344 are based on the detection of $<3$ QPEs and that further monitoring is needed.

\vspace{-0.1in}
\subsection{A crest of narrow flares}
\label{sec:crest}
\vspace{-0.05in}

Even though some QPEs show deviations from quasi-periodicity (see \citealp{giustini2024,arcodia2022,chakraborty2025, hernandezgarcia2025}), the narrow flares observed in J2344 are unprecedented. They span durations between 5 and 30 minutes, which is comparable to their separation within the same event (see Table~\ref{tab:time}). Assuming blackbody-like thermal emission, their temperature is higher than that of their corresponding QPE, although the radii of the emitting region are generally smaller (see Appendix~\ref{a:xrspec}). We note that the number of narrow flares is different between E-1 and E-2.

Within the XMM-Newton light curve, the narrow flares predominantly appear in concurrence with the broad flares, as if they were a crest to the QPEs. Moreover, they are only detected in the rising phases of such broad flares, and potentially  have similar onset times. Two dimmer flares, however, also appear isolated with respect to the broad flares (F-1 and F-2). While the XMM-Newton observations may suggest that F-1 and F-2 are stochastic features of the quiescence, the examination of the combined EP and XMM-Newton datasets suggests that this could be a recurrent feature (see Appendix \ref{a:complc}).

To gain more insights about their nature, we performed time-resolved spectral analysis of the three stacked narrow flares in E-2 (see details in Appendix \ref{a:xrspec}). As shown in Fig. \ref{fig:hysteresis_m}, we reveal a tentative clockwise rotation within the L$_{\rm bol}$ versus T plane, and a general colder-when-brighter behavior, opposite to that of QPEs. 

We note that \citealp{arcodia2022} showed that the QPEs in eRO-QPE1 would sometimes manifest as the superposition of different flares.  However, the timing differences between overlapping eruptions of eRO-QPE1 are not as extreme as the difference between the narrow and broad flares of J2344. Moreover, the spectral evolution of the individual eRO-QPE1 eruptions does not resemble that of the narrow flares of J2344. Therefore, none of the aforementioned features, or the presence of narrow flares altogether, has ever been observed in the context of QPEs. To our knowledge, no current theoretical model of QPEs describes or predicts such a crest of narrow flares. 

\vspace{-0.12in}
\section{Implications and conclusions}

The QPEs discovered in J2344 five years after its TDE-like flare corroborate the observational and theoretical association between these two phenomena. Moreover, \cite{homan2023} interpreted J2344 as a TDE in a faded active galactic nucleus (AGN). This supports the model of \cite{Jiang2025}, which suggests that previous AGN activity can boost both TDEs and EMRI rates, and therefore the emergence of QPEs. 

However, the QPEs discovered in J2344 are accompanied by crests of narrow flares, which are unprecedented in other sources of this class. Although it is hard to reconcile these observations within the current theoretical framework without invoking exotic prescriptions, here we speculate on potential explanations.

In an EMRI framework, the narrow flares could be due to clumpy debris stripped from the perturbing star at each disk passage. The works of \citealp{yao2025}, \citealp{linial2025}, and \citealp{mummery2025} have shown that such stellar debris significantly contributes to the general emission process. In particular, \cite{linial2025} have shown that for orbital periods longer than $\sim10$ h, it is the shocked material stripped from the perturbing star that dominates the emission. Conversely, for periods shorter than $\sim10\,$h, it is the ejected disk material that is responsible for the QPE emission. Since the timing of the QPEs in J2344 lies at the transition between these two regimes, the emission should be a combination of both processes. While the disk ejecta are responsible for the broad flares, we propose that the stellar debris could form clumps that give rise to the crest of discrete narrow flares. 

Alternatively, the pertuber could be an intermediate-mass black hole (IMBH). \citealp{lam2025} showed that an IMBH of $10^4 M_\odot$ will accrete material from the disk as it plunges through it. The accretion is expected to occur in a super-Eddington regime, at luminosities comparable to those of the narrow flares in J2344. We propose that this accretion process could give rise to the narrow flares. The close separation between the IMBH and the SMBH in J2344, however, implies a gravitational wave inspiral time of 1--80 yrs (using separations of 12--32 R$_g$ and \citealp{peters1964gravitational} formalism).  This would render the detection of this event quite exceptional and, potentially, even unlikely.

In both scenarios, however, it is unclear why the narrow flares exhibit a colder-when-brighter behavior. This has been observed in the variability of AGN, where the X-rays are produced by hot and/or warm coronae (e.g., \citealp{markowitz2004, ursini2020}). If the system in J2344 were in the process of forming a corona, as potentially suggested by the quiescence spectrum (see Appendix \ref{a:xrspec}), we speculate that the perturbations causing the QPEs could reverberate and cause fast oscillations in the properties of the warm corona, giving rise to the narrow flares. However, the details of this mechanism are outside the scope of this work.  

For all scenarios, a thorough theoretical exploration is necessary and should be compared to the reported properties of the narrow flares. It should also be noted that, at least in the case of J2344, a pure EMRI framework might not be sufficient to explain the complex QPE light curve, and either different or additional physical ingredients might need to be invoked. While sophisticated tests of current models are being performed through timing experiments (e.g., \citealp{arcodia2024a,chakraborty2025timing}), and SED modeling (e.g., \citealp{Guolo2025}), the peculiar features of J2344 represent a new independent test of QPE models.

Although the results presented in this letter already reveal a puzzling and unprecedented phenomenology of QPEs, the current datasets are not sufficient to fully characterize the temporal and spectral properties of the flares in J2344. Plus, previous XMM-Newton data of J2344 do not cover a baseline long enough to discuss the early presence of QPEs (further complicated by the X-ray variability reported in \citealp{homan2023}, \citealp{Malyali2025}). We emphasize the need for more data to constrain the correlation between the narrow and broad flares, 
to constrain the timing properties of the system accurately, and to derive high-statistics phase-resolved spectra of the narrow flares.

\vspace{-0.1in}

\bibliographystyle{aa} 
\bibliography{J2344.bib}

\FloatBarrier

\begin{appendix}

\section{Acknowledgements}

\begin{acknowledgements}
The authors thank the referee, Andrew Mummery, for the insightful feedback, which undoubtedly contributed to the improvement of this manuscript. P.B. is very thankful for the discussions with María José Bustamante-Rosell, Luca Broggi, and Joheen Chakraborty. This work is based on data obtained with
the Einstein Probe, a space mission supported by the Strategic Priority Program
on Space Science of the Chinese Academy of Sciences, in collaboration with
ESA, MPE and CNES (Grant No. XDA15310000, No. XDA15052100). 
GM acknowledges support from grants n. PID2020-115325GB-C31 and n. PID2023-147338NB-C21 funded by MICIU/AEI/10.13039/501100011033 and ERDF/EU. 
MG is funded by Spanish MICIU/AEI/10.13039/501100011033 and ERDF/EU grant PID2023-147338NB-C21.

R.A. was supported by NASA through the NASA Hubble Fellowship grant \#HST-HF2-51499.001-A awarded by the Space Telescope Science Institute, which is operated by the Association of Universities for Research in Astronomy, Incorporated, under NASA contract NAS5-26555.

T.L. acknowledges funding from
the EU HORIZON-MSCA-2023-DN Project 101168906 ‘TALES: Time-domain Analysis to study the Life-cycle and Evolution of Supermassive black holes’.

B.T. acknowledges support from the European Research Council (ERC) under the European Union’s Horizon 2020 research and innovation program (grant agreement No. 950533), and from the Excellence Cluster ORIGINS, which is funded by the Deutsche Forschungsgemeinschaft (DFG; German Research
Foundation) under Germany’s Excellence Strategy—EXC 2094—390783311. 

\end{acknowledgements}
\FloatBarrier

\section{XMM-Newton light curve analysis} 
\label{a:time}

Table~\ref{tab:log} shows the log of the X-ray observations used in this work. Here we discuss the timing analysis of the XMM-Newton light curve, while the Einstein Probe data is discussed in Appendix \ref{a:complc}.

We extracted the XMM-Newton pn source and background light curves between 0.2 and 10 keV using standard SAS v.21.0.0 procedures. To exclude possible instrumental effects that could mimic variabilities, we carefully checked the EPIC-pn housekeeping and auxiliary data. Exposure losses due to onboard rejection of high-energy particles were fully accounted for, as well as a few-second-long telemetry gaps. The two time jumps of the full-second counter were also corrected for during event file creation. The six counting mode intervals in the target quadrant only happened later than 26.16 hr after exposure start. No increase in event buffer overflows was observed during the flare periods. 

After rebinning the light curve to 200\,s long bins and visually inspecting it, we modeled the different flares with a combination of Gaussian functions in addition to a constant level by using the \texttt{lmfit} Python package. We identified different complexes of Gaussian profiles as E-1, E-2, and E-3, and F-1 and F-2. The results of the fitting procedure are reported in Table \ref{tab:time}. E-3 is consistent with one broad and one narrow component, although the broad component is virtually unconstrained as we only include counts up to hour 24.3;  at later times the background started to dominate the observations. We note that 
The excision of counts after hour 24.3 is very conservative, as the task \textsc{epiclccorr} of the SAS software already corrects for background flares. 

We also extracted the XMM Optical Monitor light curve of J2344, but found no evidence of variability, in agreement with the results found for other QPEs.

\begin{table}[h]
\caption{X-ray observation log of J2344.}
\begin{tabular}{llll}
\toprule
Date       & Telescope   & ObsID                                                                     & Exp. {[}ks{]} \\ \midrule
2019-12-12           & EP/FXT &       11908752146-151                                                &     7        \\
2020-06-13           & EP/FXT &       08500000356                                                              &     11.7         \\ 
2020-12-15           & XMM-Newton &    0923600201                                                                 &    99.2         \\
2021-02-02           & EP/FXT  &     08500000373  & 24        
\\ \midrule
\end{tabular}

\label{tab:log}
\end{table}

\begin{table}[]
\caption{Gaussian modeling of the XMM-Newton light curve}
\label{tab:time}
\begin{tabular}{lcccc}
\toprule
\toprule
        & Comp. & Centroid  & Amplitude    & Sigma                  \\ 
        &  &  [h]  & [cts/s]      & [h]           \\ 
\midrule
E-1     & B     & $0.92 \pm 0.11$    & $0.14 \pm 0.03$    &  $0.412 \pm 0.084$   \\
        & N-1  & $0.19 \pm 0.03$    & $0.12 \pm 0.10$      &  $0.04 \pm 0.03$     \\
        & N-2  & $0.68 \pm 0.02$    & $0.26 \pm 0.08$    &  $0.085 \pm 0.023$   \\                    
\midrule
F-1     & N    & $8.38 \pm 0.09$    & $0.05 \pm 0.02$    &  $0.096 \pm 0.081$   \\
\midrule
F-2     & N    & $8.82 \pm 0.04$    & $0.14 \pm 0.05$    &  $0.094 \pm 0.040$   \\
\midrule
E-2     & B    & $12.47 \pm 0.07$   & $0.17 \pm 0.03$    &  $0.425 \pm 0.074$   \\
        & N-1  & $11.8 \pm 0.01$    & $0.76 \pm 0.33$      &  $0.022 \pm 0.014$   \\
        & N-2  & $12.08 \pm 0.01$   & $0.60 \pm 0.07$    &  $0.055 \pm 0.004$   \\
        & N-3  & $12.4 \pm 0.01$    & $0.32 \pm 0.08$    &  $0.037 \pm 0.008$   \\
\midrule
E-3     & B    & $24.5 \pm 1.02$    & $0.20 \pm 0.20$      &  $0.301 \pm 0.301$   \\
        & N    & $24.06 \pm 0.02$   & $0.20 \pm 0.16$      &  $0.034 \pm 0.026$   \\                   
\bottomrule
\bottomrule
\end{tabular}
\end{table}

\FloatBarrier
\section{X-ray spectral analysis}

\label{a:xrspec}

We reduced the XMM-Newton pn event files using standard SAS v.21.0.0 procedures. We extract source spectra from a circular region with radius of 20" centered on the X-ray position of J2344, while background spectra are extracted from a source-free neighboring circular region with radius of 60".

\begin{figure}
    \centering
    \includegraphics[width=\columnwidth]{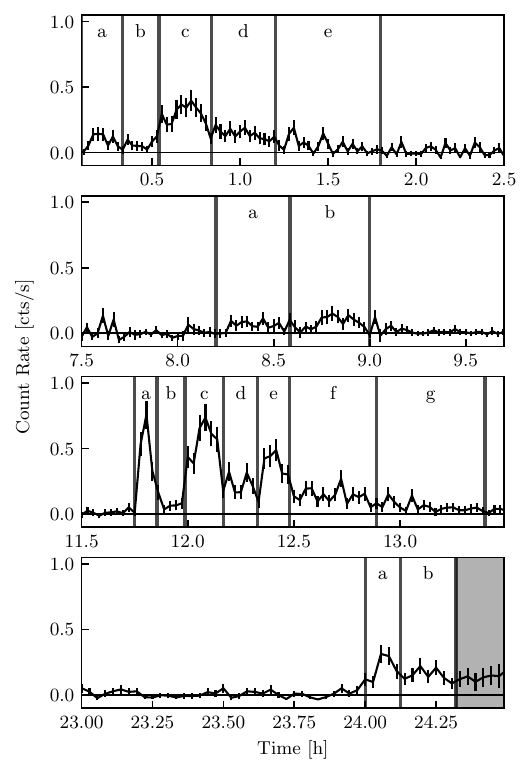}
    \caption{Different time bins extracted for the XMM-Newton pn light curve. The different panels represent (from top to bottom) the time bins for E-1, F-1 and F-2, E-2, and E-3. The results of the spectral analysis can be found in Table \ref{tab:spec}}
    \label{fig:timechunks}
\end{figure}

\begin{figure}
    \centering
    \includegraphics[width=0.9\columnwidth]{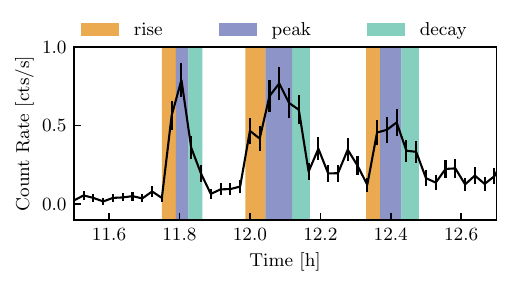}
    \caption{Different time bins extracted in order to model the rise peak and decay phases of the narrow flares in E-2. Areas corresponding to the same color were extracted and analyzed jointly.}
    \label{fig:timechunks_narrow}
\end{figure}

We first extracted source and background spectra for the quiescence level by filtering out the E-1, E-2, E-3, F-1, and F-2 flares. We then extracted spectra to model the broad and narrow features separately. The different spectral extraction bins are shown in detail in Fig. \ref{fig:timechunks}. In order to explore the rise-to-decay spectral evolution of the narrow flares, we also extracted three spectra combining the rise, peak and decay phase of the three narrow flares of E-2. This is shown in Fig. \ref{fig:timechunks_narrow}, and the results of the analysis are summarized in Fig. \ref{fig:hysteresis_m}.

All spectra were analyzed with the Bayesian X-ray Analysis software (\texttt{BXA}) version 4.1.2 (\citealp{buchner2014bxa}), which connects the nested sampling algorithm UltraNest (\citealp{buchner2019ultranest, buchner2021ultranest}) with the fitting environment CIAO/Sherpa (\citealp{fruscione2006sherpa}). Spectra were fit unbinned and using C-statistic. The fitting procedure included a PCA-based background model derived from a large sample of XMM-Newton background spectra (\citealp{simmonds2018}).

\begin{figure*}[t]
    \centering
    \includegraphics[width=0.9\linewidth]{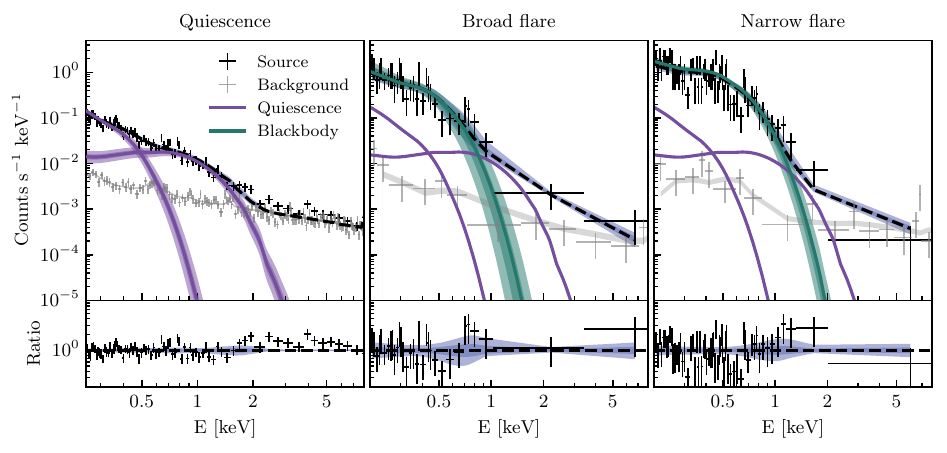}
    \caption{Example spectra and residuals of three states of J2344. The first panel shows the quiescence spectrum, the second a spectrum extracted in a phase where only the broad flare is detected (phase $b$ of E-2), and the third panel shows the spectrum of the second flare in E-2 (phase $c$). The purple lines indicate the quiescence double-blackbody model. In the first panel, we also show the 1 and 2 $\sigma$ uncertainty bands. The green line (and respective uncertainty bands) represents the additional blackbody used to model the flares. The dashed black line and blue uncertainty bands represent the total model, which also includes the background (gray band). The black and gray markers are, respectively, source and background spectral bins.}
    \label{fig:xrayspec}
\end{figure*}

We tried modeling the quiescence as a single blackbody, a multicolor blackbody (diskbb in \textsc{xspec} language) or a single powerlaw, all corrected for redshift, Galactic absorption, and in addition to the background model. However, none of these models returned satisfying residuals. We found that a double blackbody model best fits the quiescence spectrum, in agreement with the results of \citealp{homan2023} (see the first panel of Fig. \ref{fig:xrayspec}). The two blackbodies have temperatures of respectively $kT_1 = 0.234^{+0.011}_{-0.010}~\mathrm{keV}$ and $kT_2 = 0.056^{+0.003}_{-0.002}~\mathrm{keV}$. This corresponds to a 0.2-10 keV luminosity of $L_{\mathrm{qui},0.2-10}=5.4^{+0.6}_{-0.5} \times 10^{42} $ erg/s. The temperature of the hottest of the two blackbodies has increased, compared to the results of \citealp{homan2023}. The quiescence spectrum is similar to that of RXJ1301, as reported in \citealp{giustini2024} and, as discussed by the authors, it could be linked to the development of a warm corona. 

Since the quiescent model is well constrained, we freeze it when modeling both narrow and broad flares. This is based on the assumption that the process responsible for producing the flares is superimposed on the quiescent emission (i.e., the TDE disk) and does not significantly affect it. As for all other QPEs, we find that a single blackbody in addition to the quiescent emission can reproduce well the flaring spectra (see Fig. \ref{fig:xrayspec}). The results of the modeling of the different spectra are reported in Table \ref{tab:spec}. We note that the spectra of the narrow flares generally exhibit residuals around 0.5-0.8 keV, a feature also observed for example in  GSN069 (\citealp{miniutti2023, kosec2025}). However, it is not possible to further analyze these residuals due to the low spectral counts. 

Even if the spectra of the narrow flares can be well modeled with a single blackbody in addition to the quiescence model, the light curve decomposition and the interpretation of the narrow flares as superimposed to the QPEs would require them to be modeled as an additional component on top of the broad flare model. Therefore, we also attempted to model the narrow flares in E-1, E-2, and E-3 by adding two blackbodies on top of the quiescent model, respectively accounting for the underlying QPEs and the narrow flare itself. We find that the two blackbodies are highly degenerate and that the addition of a second component is not statistically justified. We therefore constrained the first blackbody to only explore the temperature and normalization space measured from the spectral analysis of the spectra of the broad flares. With such constraints, it is possible to measure the temperatures of the narrow-only flares, although with large uncertainties. The results of this analysis are reported in Table \ref{tab:spec_complex}. In Fig. \ref{fig:hysteresis_m}, we show the evolution of the broad flare in E-2, which is consistent with the classical hysteresis cycle of QPEs, although the first bin (phase $b$ of E-2) is likely affected by contamination from the narrow flares. 

\begin{table}[b]
\centering
\caption{Results of the single-blackbody time-resolved spectral modeling of the XMM-Newton dataset.}
\label{tab:spec}
\begin{tabular}{lccccc}
\toprule
\toprule
        & Comp. & $\Delta$t  & kT$_1$    & L$_{bol,1}$ & $\mathrm{R_{BB,1}}$   \\ 
        &  &         [s]  &    [eV]      & [$10^{43}$ erg/s]  & [$10^{11}$ cm]      \\ 
\midrule
E-1     & a & 414 & $50^{+7}_{-6}$ & $2.18^{+1.15}_{-0.76}$ & $5.1\pm1.8$     \\
        & b & 378 & $44^{+12}_{-9}$ & $2.56^{+3.31}_{-1.57}$ & $7.4\pm 5$   \\
        & c & 540 & $68^{+5}_{-4}$ & $3.18^{0.48}_{0.42}$ & $3.4\pm0.5$     \\ 
        & d & 648 & $57 \pm 4$ &  $2.61^{+0.62}_{-0.51}$ & $4.3 \pm 0.8$ \\
        & e & 1080 & $44^{+6}_{-5}$ &  $1.86^{+1}_{-0.68}$ & $6.1 \pm 2.1$ \\

\midrule

F-1     & a & 693 & $ 68^{+11}_{-10}$ &  $0.61^{+0.25}_{-0.18}$ &  $1.5 \pm 0.5$  \\
\midrule
F-2     & b & 747 & $ 58^{+7}_{-6}$ &  $1.45^{+0.47}_{-0.36}$ &  $3.2 \pm 0.8$   \\
\midrule

E-2     & a   & 189 &   $146^{+13}_{-1}$ & $2.18^{0.23}_{0.21}$ & $0.6\pm0.1$     \\
        & b   & 237 & $80^{+23}_{-17}$ &  $0.55^{+0.32}_{-0.21}$ & $1 \pm 0.6$\\
        & c   & 329 &  $95^{+5}_{-5}$ & $3.61^{0.33}_{0.31}$ & $1.8\pm0.2$    \\
        & d   & 306 & $71\pm7$ & $2.39^{+0.54}_{-0.47}$ & $2.7 \pm 0.6$\\
        & e  & 252 & $84\pm5$ & $3^{0.43}_{0.39}$ & $2.1\pm0.3$     \\
        & f   & 738 & $64 \pm 4$ & $1.31^{+0.23}_{-0.21}$ & $2.5 \pm 0.4$\\
        & g   & 739 & $56^{+7}_{-6}$ & $1.01^{+0.44}_{-0.3}$ & $2.8 \pm 0.8$\\
\midrule

E-3     & a  &  110  &  $102^{+12}_{-10}$ & $1.25\pm0.24$ & $0.9 \pm 0.2$\\
        & b  &  356 &  $109^{+30}_{-23}$ & $0.81\pm0.20$ & $0.7 \pm 0.3$ \\                   
        \bottomrule
        \bottomrule
\end{tabular}
\end{table}

\begin{table*}[t]
\centering
\caption{Results of the double-blackbody time-resolved spectral modeling of the XMM-Newton dataset.}
\label{tab:spec_complex}
\begin{tabular}{lcccccccc}
\toprule
\toprule
        & Comp. & $\Delta$t  & kT$_1$    & L$_{bol,1}$ & $\mathrm{R_{BB,1}}$ &  kT$_2$    & L$_{bol,2}$ &$\mathrm{R_{BB,2}}$   \\ 
        &  &         [s]  &    [eV]      & [$10^{43}$ erg/s]  & [$10^{11}$ cm]  &   [eV]      & [$10^{43}$ erg/s]  & [$10^{11}$ cm]          \\ 
\midrule
E-1     & a & 414 & & & &  $92\pm 92$ & $0.03^{+1}_{-0.03}$ & $0.2\pm0.2$    \\
        & b & 378 & $44^{+12}_{-9}$ & $2.56^{+3.31}_{-1.57}$ & $7.4\pm 5$   \\
        & c & 540 & & & &  $60^{+12}_{-28}$ & $3.66^{+8.04}_{-2.42}$ & $4.7\pm4.5$  \\ 
        & d & 648 & $57 \pm 4$ &  $2.61^{+0.62}_{-0.51}$ & $4.3 \pm 0.8$ \\
        & e & 1080 & $44^{+6}_{-5}$ &  $1.86^{+1}_{-0.68}$ & $6.1 \pm 2.1$ \\

\midrule

F-1     & a & 693 & $ 68^{+11}_{-10}$ &  $0.61^{+0.25}_{-0.18}$ &  $1.5 \pm 0.5$  \\
\midrule
F-2     & b & 747 & $ 58^{+7}_{-6}$ &  $1.45^{+0.47}_{-0.36}$ &  $3.2 \pm 0.8$   \\
\midrule

E-2     & a   & 189 & & & &  $208^{+25}_{-23}$ & $1.19^{+0.24}_{-0.2}$ & $0.2\pm0.1$     \\
        & b   & 237 & $80^{+23}_{-17}$ &  $0.55^{+0.32}_{-0.21}$ & $1 \pm 0.6$\\
        & c   & 329 & & & &  $110^{+8}_{-7}$ & $2.21^{+0.35}_{-0.29}$ & $1.1\pm0.2$    \\
        & d   & 306 & $71\pm7$ & $2.39^{+0.54}_{-0.47}$ & $2.7 \pm 0.6$\\
        & e  & 252 & & & &  $98^{+13}_{-11}$ & $1.41^{+0.45}_{-0.35}$ & $1.1\pm0.3$     \\
        & f   & 738 & $64 \pm 4$ & $1.31^{+0.23}_{-0.21}$ & $2.5 \pm 0.4$\\
        & g   & 739 & $56^{+7}_{-6}$ & $1.01^{+0.44}_{-0.3}$ & $2.8 \pm 0.8$\\
\midrule

E-3     & a  &  110  &  &  &  & $381^{+688}_{-223}$ & $0.2^{+0.3}_{-0.1}$ & $<0.1$ \\
        & b  &  356 &  $105^{+12}_{-10}$ & $0.99\pm0.13$ & $0.8 \pm 0.2$ \\                   
        \bottomrule
        \bottomrule

\end{tabular}

\tablefoot{The second blackbody is used to model the narrow flares. In coincidence with the narrow flares, the blackbody accounting for the broad flare is constrained to vary between the minima and maxima parameter values derived for the bare broad flares.
It can be seen by the large uncertainties that certain epochs reject the addition of the second blackbody component (e.g., E-1 $a$)}
\end{table*}

Lastly, in order to probe the spectral evolution along a single narrow flare, we modeled the spectra extracted as shown in Fig. \ref{fig:timechunks_narrow} once again with a blackbody on top of the quiescent emission. The results of the analysis are plotted in Fig. \ref{fig:hysteresis_m}. It can be seen that the spectral evolution tentatively follows a clockwise evolution in the L$_{ \rm bol}$ versus T plane, which differs from that of QPEs. However, even if the direction cannot be confidently identified due to large uncertainties, a clear colder-when-brighter behavior can be observed, which is also uncommon in QPEs or TDEs. 

\FloatBarrier
\noindent

\section{Comparison of Einstein Probe and XMM-Newton light curves}
\label{a:complc}

\begin{figure*}[t]
    \centering
    \includegraphics[width=0.91\linewidth]{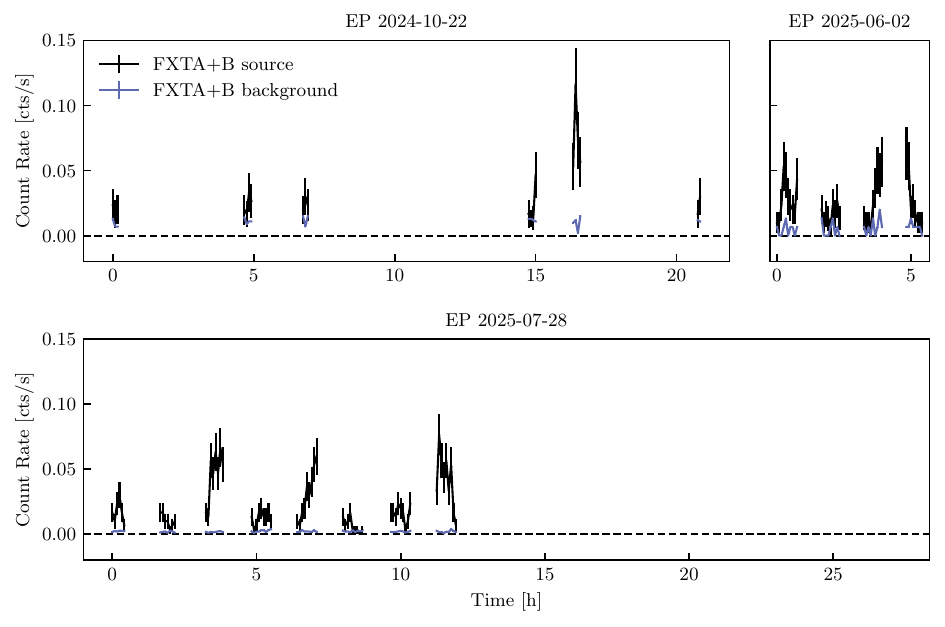}
    \caption{All Einstein Probe light curves of J2344. All times are relative to the start of each observations.}
    \label{fig:d1}
    \includegraphics[width=0.91\linewidth]{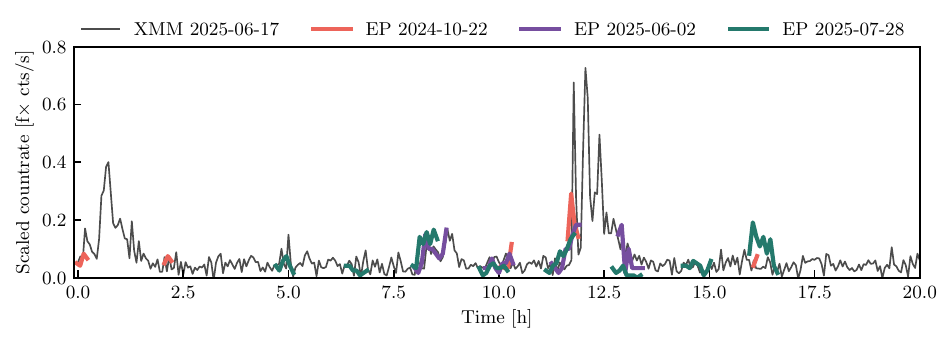}
    \caption{Comparison of the XMM newton and EP light curves. The X-axis refers to time since the start of the XMM-Newton observations. The EP light curves are rescaled by a factor 2.5 in order for the y-axes to be comparable across different instruments. The green curve shows at least one flare that cannot be matched to features observed in the XMM-Newton light curve.}
    \label{fig:eplcs}
\end{figure*}

We reprocessed all Einstein Probe data using standard recipes with the FXT Data Analysis Software Package (FXTDAS) v1.10. Source light curves were extracted between 0.2 and 5 keV with the HEASOFT \textsc{XSELECT} task from circular regions with 30" radii, while background light curves were extracted from neighboring source-free circular regions of 360" radii. The FXT-A and FXT-B light curves are binned to 300s bins and then summed. We show all the EP light curves in Fig. \ref{fig:d1}. In Fig. \ref{fig:eplcs}, we show all the X-ray light curves presented in this work in a single panel. For visualization purposes, we rescaled the countrates of the EP/FXT light curves by a factor 2.5, as inferred by comparing the nominal effective areas of FXT-A, FXT-B, and the XMM-Newton EPIC-pn at 1 keV. We shifted the EP light curves along the time axis to match the XMM-Newton observations, assuming quasi-periodicity. We find that the EP 2024 October 22 and EP 2025 June 02 observations can be well matched to the XMM-Newton light curve. In particular, the EP 2025 June 02 observations suggest that the F-1 and F-2 narrow features are periodic. Instead, it is not possible to match the latest EP observations (EP 2025 July 28) to the XMM light curve, without implying that a previously unobserved flare has now appeared at hour $\sim$16. While the first two EP light curves suggest that the narrow flares are recurrent events, the July EP light curve instead suggests that the narrow flares could also have a stochastic nature. Further monitoring of this source is necessary to draw a conclusive picture.

\newpage
\end{appendix}

\end{document}